\documentclass[doublecol]{epl2} 

\usepackage{amsmath,amssymb,amsfonts}
\usepackage{float}
\usepackage{rotating}
\usepackage{amssymb}
\usepackage{amsfonts}
\usepackage{amsmath}
\usepackage{color}
\usepackage[toc,page]{appendix}

\newcommand{\fref}[1]{Fig. \ref{#1}}

\newcommand{\eref}[1]{Eq.(\ref{#1})}
\newcommand{\erefs}[1]{Eqs.(\ref{#1})}
\newcommand{\reff}[1]{(\ref{#1})}
\newcommand{\citeref}[1]{Ref.\!\!\cite{#1}}
\newcommand{\citerefs}[1]{Refs.\!\!\cite{#1}}

\newcommand{\p}{\partial}
\newcommand{\ie}{\emph{i.e., }}
\newcommand{\oms}{\omega^{*}}
\newcommand{\omus}{\omega_{1}^{*}}
\newcommand{\pvar}{\bar{k}}
\newcommand{\viscd}{\eta_{v}}
\newcommand{\visc}{\nu}
\newcommand{\viscbar}{\bar{\nu}}

\title{Study of MRI in Stratified Viscous Plasma Configuration}
\shorttitle{MRI in Stratified Plasma Disks} 

\author{Nakia Carlevaro\inst{1,2} \and \and Giovanni Montani \inst{1,3} Fabrizio Renzi\inst{3}}
\shortauthor{N. Carlevaro, G. Montani, F. Renzi}

\institute{
  \inst{1} ENEA, Fusion and Nuclear Safety Department, C.R. Frascati, Via E. Fermi, 45 (00044) Frascati (RM), Italy;\\
  \inst{2} L.T. Calcoli, Via Bergamo, 60 (23807) Merate (LC), Italy;\\
  \inst{3} Department of Physics, ``Sapienza'' University of Rome, P.le Aldo Moro, 5 (00185) Roma, Italy.
}

\abstract{We analyze the morphology of the Magneto-rotational Instability (MRI) for a stratified viscous plasma disk configuration in differential rotation, taking into account the so-called corotation theorem for the background profile. In order to select the intrinsic Alfv\'enic nature of MRI, we deal with an incompressible plasma and we adopt a formulation of the local perturbation analysis based on the use of the magnetic flux function as a dynamical variable. Our study outlines, as consequence of the corotation condition, a marked asymmetry of the MRI with respect to the equatorial plane, particularly evident in a complete damping of the instability over a positive critical height on the equatorial plane. We also emphasize how such a feature is already present (although less pronounced) even in the ideal case, restoring a dependence of the MRI on the stratified morphology of the gravitational field.}

\begin{document}

\maketitle

\section{Introduction}
In 1959, E.P. Velikhov discovered a new type of magneto-hydrodynamics (MHD) instability, associated to the coupling of the plasma differential rotation with the Alfv\'enic modes \cite{Ve59}. In this original work, it is investigated the behavior of a magnetized plasma, lying between two rotating cylinders at different angular velocity and the corresponding unstable mode spectrum was named Magneto-Rotational Instability (MRI). In 1960, S. Chandrasekar proposed the implementation of MRI to the astrophysical context, with reference to a stellar accretion disk rotating mainly due to the gravity of the central body \cite{Chand}. Then, in 1991, MRI was re-analyzed \cite{B&H1991} with the awareness of the role it can play in rising turbulence in accretion disks and, hence, the required effective viscosity to account for the Shakura idea of accretion \cite{Sh73,SS73} (for a detailed discussion of MRI in accreting structures and its contribution to the angular momentum transport, see \citeref{balbus03} and references therein).

Since MRI is an Alfv\'enic mode, it is not associated to matter transport and it survives also in incompressible plasmas, where its real nature is indeed well-traced. For a study of MRI in the case of a liquid magnetized metal, see \citeref{JGK}, and a similar astrophysical study can be found in \citeref{sano99}. In the present paper, we concentrate on such a restricted case of an incompressible plasma in order to select a pure Alfv\'enic instability and evaluate the role the so-called corotation theorem \cite{Ferraro} plays in the MRI features. Such a theorem states that the differential rotation profile of a steady axisymmetric disk (actually the background configuration on which our perturbation analysis is performed) must depend on the magnetic field morphology via the magnetic flux function only. In \citerefs{mp13,mp15,mcp16}, the same question has been addressed in the context of ideal MHD mainly focusing on parallel propagating perturbations, while here we address the behavior of a viscous incompressible rotating plasma. A valuable discussion of MRI in the case of a viscous-resistive plasma disk (but without retaining the corotation theorem and the stratified nature of the disk) has been provided in \citeref{Shak&Post}. In what follows, we will neglect (differently from \citeref{Shak&Post}) the effect of dissipation on the background profile, but we deal with a stratified disk for which the angular velocity depends on the vertical coordinate too \cite{B1995}.

The main merit of the present analysis is to outline the dependence of the MRI growth rate on the vertical profile of the background and, in particular, we show a clear asymmetry with respect to the equatorial plane. In fact, MRI is suppressed over a critical height on the equatorial plane (in the region of positive vertical cylindrical coordinate $z$, when the magnetic field has a natural dipole like configuration), by a mechanism very similar to the one investigated in \citeref{Shak&Post} as an effect of the increasing value of the viscosity coefficient. Here, the same dimensionless parameter adopted in the discussion of that article varies with $z$ as a consequence of the stratified structure of the disk, \ie the magnetic field morphology depends on the height on the equatorial plane. Since such dimensionless viscosity parameter results to increase with $z$, we observe a significant damping of the MRI growth rate. However, we emphasize how such a damping is the combined effect of what clarified above, together with an ideal property of the MRI emerging only when the corotation theorem is considered. In fact, a critical $z$ value, over which MRI is suppressed, exists already in the ideal case and it remains the same for the viscous plasma, but the vertical damping is much more marked for that case.

Also the critical wavenumber at which MRI is removed appears to be the same both in the inviscid and viscous case, but in the latter case, the suppression is evidently well marked before the critical wave number, since the growth rate rapidly decays. We also show how the ideal dispersion relation can be consistently derived from the analysis in \citeref{B1995}, as soon as the corotation constraint is taken into account. However, also in the ideal case, we are able to trace the dependence of the MRI growth rate on the magnetic field profile, a feature due just to the corotation theorem and, indeed, absent in \citeref{B1995}.

With respect to the study in \citeref{Shak&Post}, in addition to the co-rotation theorem, we also generalize the perturbation scheme to a stratified configuration. For a purely vertical magnetic field and a disk angular frequency depending on the radial variable only (as assumed in \citeref{Shak&Post}) our dispersion relation overlaps the one derived in \citeref{Shak&Post}. This is due to the fact that, under such hypotheses, the corotation theorem is automatically fulfilled.

For what concerns studies which demonstrate the relevance of the vertical matter distribution in the disk toward the efficiency of MRI in generating turbulence, see \citeref{gressel13}, where the relevance of the boundary condition is outlined, and \citeref{bodo12,brande95}. Moreover, an interesting analysis concerning an effect of saturation for MRI due to the presence of magnetosonic waves, which also influences the emergence of a turbulent regime, is provided in \citeref{live12,kno05,umu07,live09}.

Summarizing, the present analysis provides a clear picture of how the coupling of the corotation condition for the background and the stratified nature of the configuration can alter the expected morphology of the MRI, producing an important asymmetry of the corresponding growth rate with respect to the equatorial plane of the background configuration.

\section{Basic equations}
In order to set up the fundamental formalism and notation at the ground of the perturbation analysis we are going to perform, it is worth writing down the full set of basic equations governing the dynamics of a magnetized viscous fluid. As a first step, let us consider the system composed by the Faraday equation and the generalized Ohm law:
\begin{align}
\p_t \vect{B} = - c\nabla\times\vect{E}\label{Eqt.FaradayLaw}\;,\\
\vect{E} + \vect{v}\times\vect{B}/c = 0 \label{Eqt.ElecForceBalance}\;,
\end{align}
where $\vect{E}$ and $\vect{B}$ denote the electric and magnetic field, respectively, and $\vect{v}$ is the disk velocity field. In terms of the electric potential $\Phi$ and the vector potential $\vect{A}$, \eref{Eqt.ElecForceBalance} can be rewritten as follows
\begin{equation}\label{Eqt.1}
\nabla\Phi + \p_t\vect{A}/c = \vect{v}\times\vect{B}/c\;,
\end{equation}
where $\vect{B} = \nabla\times\vect{A} $ and $\nabla\cdot\vect{A}=0$. In what follows, we consider a two-dimensional axisymmetric system (using cylindrical coordinates $(r,\,\phi,\,z)$) in which all the physical variables are independent of the azimuthal angle $\phi$. Without any loss of generality, we express the magnetic field via the magnetic flux surface $\psi$ as
\begin{equation}\label{Eqt.MagneticField}
\vect{B}\equiv-\vect{e}_r\,\p_z\psi/r+\vect{e}_\phi\,\bar{B}_\phi/r+\vect{e}_z\,\p_{r}\psi/r\;,
\end{equation}
(here $\vect{e}_{r,\phi,z}$ denotes the coordinate versors). This expression of the magnetic field is associated to a vector potential having the following form
\begin{align}
\vect{A}=\vect{A}_p+\frac{\vect{e_}\phi}{r}\,\psi\;,
\end{align}
where the poloidal vector potential $\vect{A}_p$ satisfies, in the Coulomb gauge, the conditions
\begin{align}
\nabla\times\vect{A}_p=\frac{\vect{e}_\phi}{r}\,\bar{B}_\phi\;,\qquad
\nabla\cdot\vect{A}_p = 0\;.
\end{align}
In the same way, the velocity field can be split into a poloidal and an azimuthal component defined as
\begin{align}
\vect{v} = \vect{v}_p + r\omega\vect{e}_\phi\;,\qquad
\vect{v}_p = v_r\vect{e}_r + v_z\vect{e_}z\;,
\end{align}
and we can thus separate the azimuthal and poloidal parts of \eref{Eqt.1} as
\begin{subequations}
\begin{align}
\p_t\psi + \vect{v}_p\cdot\nabla\psi=0\;,\label{Eqt.GenPsi}\\
c \nabla\Phi + \p_t\vect{A}_p=\omega\nabla\psi+\vect{v}_p\times(\nabla\times\vect{A}_p)\;,\label{Eqt.2b}
\end{align}
\end{subequations}
respectively. 
Taking the curl of \eref{Eqt.2b}, we can built up a scalar equation which governs the azimuthal magnetic field $\bar{B}_\phi$:
\begin{align}
\p_t \bar{B}_\phi+\vect{v}_p\cdot\nabla \bar{B}_\phi + \bar{B}_\phi (\p_r v_r+\p_z v_z)-2 \bar{B}_\phi v_r/r =\quad\nonumber\\
= r\left( \partial_z\omega\partial_r\psi - \partial_r\omega\partial_z\psi\right)\;.
\label{Eqt.AzimtInduction}
\end{align}
This equation coincides with the azimuthal component of the so-called induction equation, and, it is worth stressing that \eref{Eqt.GenPsi} is gauge invariant, since it corresponds to the azimuthal component of the generalized Ohm law which is intrinsically gauge independent. 

Let us now face the analysis of the momentum conservation equations. In the case of a viscous fluid, the azimuthal component of the MHD Navier-Stokes equation reads
\begin{align}
\rho r\left(\p_t\omega + \vect{v}_p\cdot\nabla\omega\right) + 2\rho v_r\omega =\qquad\qquad\qquad\qquad\quad\nonumber\\
=\frac{1}{4\pi r^2}\left(\p_r\psi\p_z \bar{B}_\phi-\p_z\psi\p_r \bar{B}_\phi\right) + \viscd\nabla^2\left(r\omega\right)\;, 
\label{Eqt.EulerAzimuthal}
\end{align}
where $ \rho $ is the mass density and $\viscd$ the dynamical viscosity of the plasma, taken here as a constant quantity. Furthermore, the poloidal component assumes the form \cite{Bisno}
\begin{align}
\rho\left(\p_t\vect{v}_p+\vect{v}_p\cdot\nabla v_p\right)-\rho r \omega^2\vect{e}_r =
\qquad\qquad\qquad\nonumber\\
=-\frac{1}{4\pi r^2}\Big[\p_r\Big(\frac{1}{r}\p_r\psi\Big)+\frac{1}{r}\p_z\psi\Big]\nabla\psi +\quad\nonumber\\
-\nabla P +\vect{F}_p - \frac{\nabla \bar{B}^2_\phi}{8\pi r^2} + \viscd\nabla^2\vect{v}_p\;,
 \label{Eqt.EulerPoloidal}
\end{align}
here $P$ is the thermostatic pressure and $\vect{F}_p $ an external force acting on the plasma and it coincides with the star gravity, being written as
\begin{align}
\vect{F}_p=-\rho\omega_k^2\vect{r}_p\;,
\end{align}
where $\vect{r}_p=r\vect{e}_r+z\vect{e}_z$ and $\omega_k=\sqrt{GM(r^2+z^2)^{-3/2}}$ is the Keplerian angular frequency ($M$ being the central object mass and $G$ the gravitational constant).

The dynamical system composed by \erefs{Eqt.GenPsi}, \reff{Eqt.AzimtInduction}, \reff{Eqt.EulerAzimuthal} and \reff{Eqt.EulerPoloidal} is completed by the continuity equation (mass conservation):
\begin{equation}
\p_t\rho + \nabla\cdot\left(\rho\vect{v}\right)=0\;.\label{Eqt.Continuity}
\end{equation}
We do not assign a specific equation of state for a plasma as a whole (relating the mass density to the pressure) discussing separately the background and perturbation cases.

\section{Linear perturbations in a viscous magnetized fluid: dispersion relation} 
The linear perturbation analysis is performed by choosing a background configuration corresponding to a purely differentially rotating (at $\omega_0$) plasma disk, \ie the background poloidal component of the velocity field vanishes so that $\vect{v}_0\equiv r\omega_0\vect{e}_\phi$. The plasma is embedded in a poloidal magnetic field, associated to the background magnetic flux function $\psi_0$ via \eref{Eqt.MagneticField} (here and in what follows, we denote by the suffix 0 the background and by the suffix 1 the corresponding linear perturbations, \ie for a generic quantity $A\to A_0+A_1$).

The main assumption we adopt in our analysis is the Alfv\'enic nature of the perturbations, thus we can neglect the contribution of the magnetosonic waves and we can assume the plasma incompressibility, \ie $\nabla\cdot\vect{v}_{p}=\nabla\cdot\vect{v}_{p1}=0$ (indeed, Alfv\'en waves do not transport matter). In the case of a purely differentially rotating plasma, \erefs{Eqt.AzimtInduction} and \reff{Eqt.EulerAzimuthal} are automatically satisfied at zeroth order, while \eref{Eqt.EulerPoloidal} splits in two background equations
\begin{align}
\nabla P_0 -\rho_0(\omega^2_0 r\vect{e}_r-\omega^2_k\vect{r}_p)=0\;,\label{dhjbdslvbob}\\
\bar{\Delta}\psi_0 \equiv \frac{1}{4\pi r}\Big[\p_r\,\Big(\frac{1}{r}\p_r\Big)+\frac{1}{r}\p_z^2\Big]\psi_0=0\;.\label{dsòoijvdèsoijv}
\end{align}
Let us now focus our attention on this background system. \eref{dsòoijvdèsoijv} is the force-free condition for the vacuum magnetic field of the central object. In what follows, we will adopt a dipole configuration for $\psi_0$ which is a natural choice for compact astrophysical objects. In fact, far enough from the center (in the disk), the magnetic field is essentially dipole-like. We moreover stress that, for a thin disk (almost coinciding with the equatorial plane), the dipole field reduces to a pure vertical one. \eref{dhjbdslvbob} describes instead the gravostatic equilibrium and determines the disk morphology: note that the background pressure gradient is not negligible in our analysis. 
From this equation, we see that the presence of vertical shear in the problem (\ie the $z$-dependence of the angular velocity) is due to the vertical pressure gradient. In what follows, we develop a local perturbation approach dealing with wavelengths smaller than the scale of background variation. As far as the vertical shear is smooth (in typical accretion disks it is of the half-depth order), the local approach almost overcome the global one \cite{PAPA,B1995}. But, when the coupling between the background vertical gradient and the perturbation is relevant (see for instance the analyses in \cite{COPPI08} and \cite{Liverts}), the prediction of the local and global approach can deviate, the latter depending significantly on the boundary conditions.

Finally, we observe that, a steady background having $B_{\phi0}=const.$ (in particular $B_{\phi0}\equiv0$) and vanishing poloidal velocities is characterized by a vanishing left-hand-side of \eref{Eqt.AzimtInduction}. Therefore, from the right-hand-side, the proportionality between the angular velocity and surface function gradients comes out, leading to $\omega_0=\omega_0(\psi_0)$: this issue corresponds to the so-called corotation theorem \cite{Ferraro}.

Let us now separate, without loss of generality, the angular velocity into its corotation and generic parts as follows
\begin{align}
\omega=\bar{\omega}(\psi)+\oms\;,
\end{align}
where, we have $\omega_{0}^{*}=0$ (since $\omega_0=\bar{\omega}_0(\psi_0)$) and
\begin{equation}
\omega_1=\bar{\omega}_1+\omus=\psi_1\frac{d\bar{\omega}}{d\psi}\Big|_{\psi_0}+\omus\equiv
\omega^{\prime}_{0}\,\psi_1+\omus\;,
\label{Eqt.Omega1definition}
\end{equation}
here, the relation $\nabla\omega_0=\omega^{\prime}_{0}\nabla\psi_0$ holds. Introducing the poloidal plasma shift $\vect{\xi}_p=\xi_r\vect{e}_r+\xi_z\vect{e}_z$ defined by $\vect{v}_{1p}\equiv\partial_t\vect{\xi}_p$ and perturbing \eref{Eqt.GenPsi}, one can now write the basic relation 
\begin{align}\label{nhiugbui}
\psi_1=-\vect{\xi}_p\cdot\nabla\psi_0\Leftrightarrow
\p_t\psi_1=-\vect{v}_{1p}\cdot\nabla\psi_0\;.
\end{align}
We now observe that, in the linear perturbation regime, the induced poloidal magnetic field remains much smaller than the background component, \ie $|\nabla\psi_1|\ll|\nabla\psi_0|$. Moreover, the behavior of the perturbed pressure $P_1$ will be determined by preserving the incompressibility along the plasma dynamics.

We now write the first order quantity as follows (preserving the axial symmetry, \ie assuming no propagation along the $\phi$ direction)
\begin{align}
A_1=\tilde{A} e^{-i(\vect{k}_p\cdot\vect{r}_p-\Omega t)}
=\tilde{A} e^{-i(k_rr + k_zz - \Omega t)}\;,
\end{align}
where $\tilde{A}$ is a small constant amplitude while $\vect{k}_p=k_r\vect{e}_r+k_z\vect{e}_z$ and $\Omega$ are the poloidal wave vector and the frequency of the perturbation, respectively.
According the local approach to the perturbation dynamics, we require that the condition $\vect{k}_p\cdot\vect{r}_p\gg1$ holds.

Using \eref{Eqt.Continuity} in its perturbed form and retaining only the leading terms in $\vect{k}_p$, we can now rewrite \erefs{Eqt.AzimtInduction}, \reff{Eqt.EulerAzimuthal} and \reff{Eqt.EulerPoloidal} as
\begin{align}
\Omega\bar{B}_{\phi1}-r^2\omus\vect{k}_p\cdot\vect{B}_0=0\;,\label{Eqt.3}\\
ir(i\Omega + \visc k_p^2)\omus-2\Omega\omega_0\xi_r+\qquad\qquad\qquad\qquad\qquad\nonumber\\
+\frac{\vect{k}_p\cdot\vect{B}_0}{4\pi r\rho_0}\bar{B}_{\phi1}+ i\visc k_p^2 r\omega_0\psi_1=0\;,\label{Eqt.4}\\
i\Omega(i\Omega + \visc k_p^2)\vect{\xi}_p-2r\omega_0(\omega^{\prime}_{0}\psi_1 + \omus)\vect{e}_r+\qquad\qquad\nonumber\\
-\frac{i\vect{k}_p P_1}{\rho_0} - \frac{k_p^2\nabla\psi_0}{4\pi r^2 \rho_0}\psi_1=0\;,\label{Eqt.5}
\end{align}
respectively, where $\visc\equiv\viscd/\rho_0$ and we have to take into account the incompressibility constraint $\vect{k}_p\cdot\vect{\xi}_p=0$. Above, we also used the relation $\omega^{\prime}_{0}\p_t\psi_1=-\omega^{\prime}_{0}\vect{v}_{1p}\cdot\nabla\psi_0\equiv-\vect{v}_{1p}\cdot\nabla\omega_0$ which is guaranteed by \eref{nhiugbui}. Analogously, in \eref{Eqt.3} the contribution $\omega^{\prime}_{0}\psi_1$ naturally cancels in the right-had side.

Combining \erefs{Eqt.3} and \reff{Eqt.4}, we get
\begin{align}
r\p_r\psi_0(-i\Omega(i\Omega+\visc k_p^2)-\omega^2_A)\omus+\nonumber\qquad\qquad\qquad\qquad\\
+2\omega_0\Omega^2\p_r\psi_0\psi_1-\visc k^2 i\Omega y_r/2\omega_0=0\;,
\label{Eqt.Perturbed_Omega2}
\end{align}
where $\omega_A^2\equiv(\vect{k}_p\cdot\vect{v}_A)^2$ is the frequency associated to the Alfv\'en speed $v_A=\sqrt{B_0^2/4\pi\rho_0}$ and $\varphi\equiv2r\omega_0\omega^{\prime}_0\p_r\psi_0 $ includes information about the angular velocity gradient. Let us now take the scalar product of \eref{Eqt.5} with the wave vector $\vect{k}_p$ to obtain the behavior of the perturbed pressure $P_1$
\begin{align}
iP_1=-\Big[2r\rho_0\omega_0(\omega^{\prime}_0\psi_1+\omus)\frac{k_r}{k_p^2}+
\frac{\vect{k}_p\cdot\nabla\psi_0}{4\pi r^2}\,\psi_1\Big]\;,
\end{align}
where we have required that the incompressibility condition is preserved along the dynamics. Substituting the expression above in \eref{Eqt.5}, multiplying it by $\nabla\psi_0$ and noting that $\psi_1=-\vect{\xi}_p\cdot\nabla\psi_0$, we obtain the following basic equation
\begin{align}
\left[-i\Omega(i\Omega + \visc k_p^2) - \omega^2_A - \delta \varphi \right]\psi_1
=2\delta r\omega_0\p_r\psi_0\omus\;,
\label{Eqt.Perturbed_Psi1}
\end{align}
where
\begin{align}
\delta\equiv 1 - \frac{k_r(\vect{k}_p\cdot\nabla\psi_0)}{k_p^2\partial_r\psi_0}\;.
\end{align}
In the same way, the radial component of \eref{Eqt.5} provides $\xi_r$ in terms of $\psi_1$ and $\omus$:
\begin{align}
-i\Omega(i\Omega+\visc k_p^2)\p_r\psi_0\xi_r =\qquad\qquad\qquad\qquad\qquad\nonumber\\
=-(\alpha \varphi + \delta k_p^2 v^2_{Az})-2\alpha r\omega_0\p_r\psi_0\omus\;,
\label{Eqt.Perturbed_Xir}
\end{align}
with
\begin{align}
\alpha=1-k^2_r/k_p^2\;,\qquad 
v_{Az}=\p_r\psi_0/\sqrt{4\pi\rho_0 r^2}\;.
\end{align}

Combining together \erefs{Eqt.Perturbed_Omega2}, \reff{Eqt.Perturbed_Psi1} and \reff{Eqt.Perturbed_Xir}, leads to the following dispersion relation
\begin{align}
q^4+\delta\left[\varphi\omega^2_A-s^2\Omega^2\right]=0\;,
\label{Eqt.Dispersion}
\end{align}
where we have defined
\begin{align}
q^2=-i\Omega(i\Omega+\visc k_p^2)-\omega^2_A\;,\quad
s^2=4\alpha\omega^2_0/\delta+\varphi\;.
\end{align}
In what follows, we extract information from such a dispersion relation to characterize the MRI validity regions.

\section{Physical implications}
For further analysis, it is convenient to rewrite \eref{Eqt.Dispersion} in a dimensionless form. Thus, we introduce the following variables:
\begin{equation}
y=\frac{i\Omega}{\omega_0}\;,\quad
\pvar=\frac{\omega_A}{\omega_0}\;,\quad
\bar{s}^2=\frac{s^2}{\omega^2_0}\;,\quad
\viscbar=\frac{\visc\omega_0}{\chi^2_A}\;,
\end{equation}
where, $\chi_A=\omega_A^2/k_p^2=\delta^2 v_{Az}^2/\alpha$ denotes an effective Alfv\'en speed and we underline how, for fixed magnetic field, the variable $\pvar$ properly represents a normalized wave vector. Using such definitions, the dispersion relation takes the form
\begin{align}\label{Eqt.Adispersion}
y^4 + 2\viscbar \pvar^2 y^3 + (2\pvar^2 + \delta\bar{s}^2 + \viscbar^2\pvar^4)y^2+\qquad\qquad\nonumber\\
+2\viscbar\pvar^4y+\pvar^2(\delta(\bar{s}^2-4\alpha/\delta)+\pvar^2)=0\;.
\end{align}
Such a quartic equation in the $y$ variable can be analytically studied only in some simplified cases \cite{B&H1998,Shak&Post,Pess&Chan}. Thus, we numerically integrate \eref{Eqt.Adispersion} focusing on the solutions in $y$ having a positive real part, corresponding to unstable modes with $\textrm{Im}[\Omega]<0$.

In order to elucidate the physical content of the dispersion relation, let us now consider $\alpha=const.$, constraining the orientation of the wavenumber $\vect{k}_p$ in the $(r,\,z)$ plane. Moreover, the stellar magnetic field can be reliably represented by a background dipole-like configuration (satisfying the force-free condition, \ie \eref{dsòoijvdèsoijv}) as
\begin{align}
\psi_0(r^2,z)=\mu_0r^2(r^2+z^2)^{-3/2}\;,
\end{align}
with $\mu_0=const.$ The gravitational field is retained as Newtonian, and we use the functional dependence of the angular velocity on the equatorial plane
$\omega_0^2|_{z=0}=\omega^2_{k}|_{z=0}=GM\mu^3_0/\psi_0^{3}|_{z=0}$, assuming to extend this expression everywhere in the disk by virtue of the corotation theorem:
\begin{equation}
\omega^2_{0}=GM\mu^3_0/\psi^3_0\;.
\end{equation}
This relation allows to predict the behavior of the rotation profile even far away from the midplane.

\subsection{The stratified ideal case}
Let us now discuss the solutions of \eref{Eqt.Adispersion} in the inviscid limit. For vanishing viscosity, the dispersion relation takes the following form
\begin{equation}\label{hjgvbkjhvh}
y^4+y^2(\delta\bar{s}^2+2\pvar^2)+\pvar^2(\delta(\bar{s}^2-4\alpha/\delta)+\pvar^2)=0\;.
\end{equation}
This equation is the same found in \citeref{B1995} for a divergent polytropic index and, in the limit of $k_r=0$ (or equivalently for vanishing radial magnetic field), it reduces to the proper dispersion relation for a thin accretion disk with a Keplerian rotation profile \cite{B&H1991,mp15}. The main difference, here, is the appearance of the factor $\delta$ which contains informations on the magnetic field behavior in the three-dimensional space through the ratio $\p_z\psi_0/\p_r\psi_0$ (once fixed the value of $\alpha$). The physical content of the inviscid dispersion relation is then summarized by the instability condition:
\begin{equation}\label{fhfdjksjdh}
\pvar<\pvar_c\equiv\sqrt{\delta(4\alpha/\delta-\bar{s}^2)}\;,
\end{equation}
while, for $\pvar\geqslant\pvar_c$, one gets $y=0$. Moreover, \eref{hjgvbkjhvh} is a simple quadratic form in $\Omega^2 $ and it is easy to show that a maximum unstable growth rate exists:
\begin{align}
y_M=\delta r\p_r\omega_0/(2\omega_0\alpha)\;,
\end{align}
occurring when the Alfv\'en frequency assumes the following expression
\begin{align}
\omega^2_{A}=\omega^2_{A(M)}\equiv=-r\beta\p_r\omega_0(\omega_0 + r\beta\p_r\omega_0/4\alpha)\;.
\end{align}
For a dipole-like configuration, which properly describes stellar magnetic fields, we get $\p_z\psi_0/\p_r\psi_0=3zr/(r^2-2z^2)$ and three different cases can be distinguished accordingly:
\begin{figure}[H]
\centering
\includegraphics[width=0.8\columnwidth]{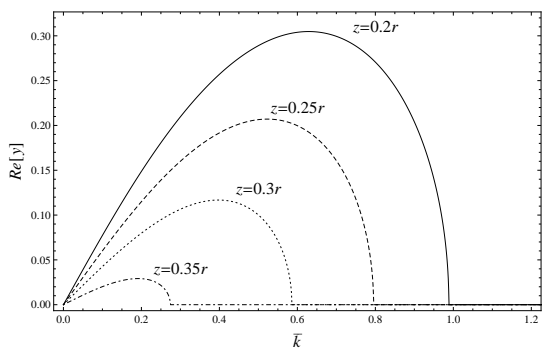}
\caption{Effect of the height on the MRI in inviscid disks with dipole-like magnetic field. The curves represent real solutions (unstable) of \eref{hjgvbkjhvh} as function of the dimensionless wave vector $\pvar$ and they are obtained fixing the radial coordinate $r$ and the disk background configuration: this leads to unstable modes which differ only for the value of $z$ (as indicated in the plot). Increasing the height, the MRI interval shifts to low wave vectors and shrinks.\label{Fig.MRIwithz}}
\end{figure} 
\begin{enumerate}
\item[\emph{(i)}]$z>0$. In this case, $\delta$ decreases for increasing value of $z$. For $\delta>0$, a critical height $z^*$ exists where $\pvar_c=0$ and the MRI is completely suppressed by virtue of the condition \reff{fhfdjksjdh}.
In fact, the standard stability constraint for a magnetized disk \cite{B1995} is found to be always satisfied for $z\geqslant z^*$. It is worth stressing that, where $\alpha<\delta<0$, the criterium \reff{fhfdjksjdh} reads $\omega_A^2<-\delta\varphi$. Clearly, this region becomes more and more stable while approaching $z^*$. In \fref{Fig.MRIwithz}, the $\delta$ suppression is reported for different heights.

\item[\emph{(ii)}]$z=0$. On the equatorial plane, $\p_z\psi_0=0$ and, thus, $\delta=\alpha$. The resulting dispersion relation is a generalization of what is found for a Keplerian disk under the assumption $\vect{k}_p\parallel \vect{B}_0 $ (which, in the flux surface formalism, reads $\vect{k}_p\cdot\nabla\psi_0=0$ \cite{mp15}). In this case, changing the orientation of the wave vector in the $(r,\,z)$ plane reproduces a suppression similar to that discussed for the case $z>0$. In \fref{Fig.MRIz0}, the midplane behavior corresponding to $\alpha=0.7$ is reported.

\item[\emph{(iii)}]$z<0$. Below the equatorial plane, $\delta>\alpha$, and the situation is the opposite of the case \emph{(i)}. The resulting MRI mode is enhanced but the amplification is weak and the midplane behavior is resembled because $\delta$ is essentially generally of order unity. Nonetheless, for peculiar magnetic configurations, it is possible to get significant enhancement for the MRI mode.
\end{enumerate}
\begin{figure}[H]
\centering
\includegraphics[width=0.8\columnwidth]{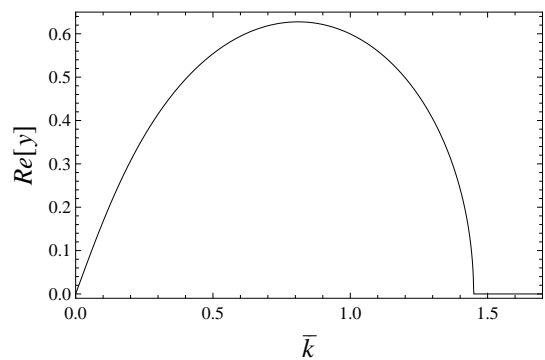}
\caption{Maximum growth rate for inviscid disks with dipole-like magnetic field. The radial coordinate is the same of \fref{Fig.MRIwithz}. The curve represents the real solution of \eref{hjgvbkjhvh} for $z=0$ as function of the dimensionless wave vector $\pvar$.\label{Fig.MRIz0}}
\end{figure}
\begin{figure}[H]
\centering
\includegraphics[width=0.8\columnwidth]{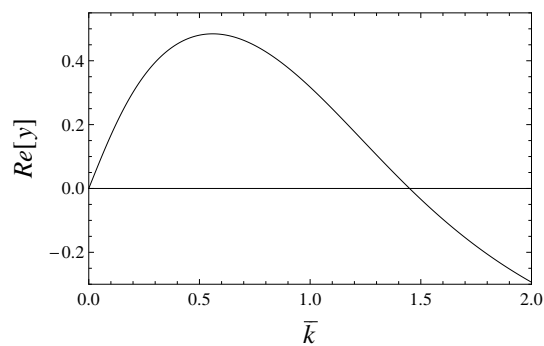}
\caption{Same as in \fref{Fig.MRIz0} for viscous disks with $\viscbar=1$. The critical point $\pvar_c$ does not vary with respect to the inviscid case, but the mode amplitude decreases by about 1/3 (cfr. \fref{Fig.MRIz0}). For $\pvar>\pvar_c$, the mode turns into a damping proportional to $\viscbar\pvar^2$.\label{Fig.MRIviscous_z0}}
\end{figure}

\subsection{The stratified viscous case}
We are now going to discuss the general solution of \eref{Eqt.Adispersion}.
It describes the stability behavior of an incompressible stratified viscous disk and, as discussed above, the system geometry enters the dispersion relation through the factor $\delta$. It is moreover easy to verify that it reduces to the proper dispersion relation for adiabatic perturbations in a thin viscous magnetized disk \cite{Shak&Post,Pess&Chan}, or equivalently to that found for a rotating metal annulus \cite{JGK}, for vanishing $k_r$. The effect of viscosity is to make more stable the disk configuration and, consequently, the MRI has a lower growth rate with respect to the inviscid case. 	

In \fref{Fig.MRIviscous_z0}, we plot the unstable solution on the equatorial plane for $\viscbar=1$ and $\alpha=0.7$. This is actually the same case of \fref{Fig.MRIz0} and, here, the effect of the viscosity can be easily recognized comparing the maximum growth rate.
\begin{figure}[H]
\centering
\includegraphics[width=0.8\columnwidth]{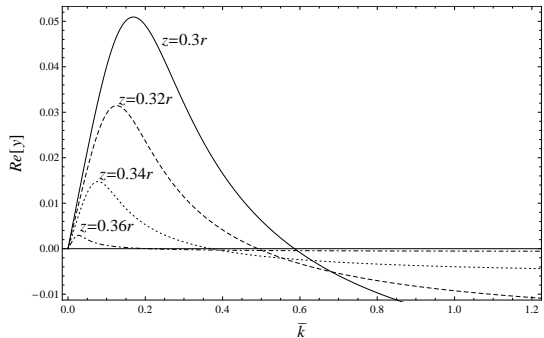}
\caption{Influence of the height on MRI in viscous disks with dipole-like magnetic field. The curves are the unstable solutions to \eref{Eqt.Adispersion}, in a narrow range of $z$ near $z^\star$. In this region, viscosity dominates the dynamics and enhances the suppression due to a stratified configuration leading to unstable modes whose features depend on $\viscbar$. The same transition is observed in a thin disk at increasing viscosity \cite{Shak&Post,Pess&Chan}. As in \fref{Fig.MRIwithz}, curves differ only in the vertical coordinate value.\label{Fig.MRIviscous_withz}}
\end{figure}

It is worth stressing that, in a stratified disk, the dimensionless viscosity parameter $\viscbar$ is also a function of the height $z$ (through the effective Alfv\'{e}n speed $\chi_A$) and, in the region where $\alpha\leqslant\delta\leqslant0$, it grows by several orders of magnitude. In fact, the magnetic field decreases in amplitude approaching the critical height $z^{*}$ and the effect of viscosity combines with the height suppression. When $\viscbar\gg1$, viscosity dominates the perturbation dynamics and the morphology of the instability changes. One can also verify that the $z$ dependence of the growth rate  (in this region of the disk) reproduces exactly the same behavior of the increase the disk viscosity at $z=0$. In \fref{Fig.MRIviscous_withz}, the unstable solution is plotted for different heights near $z^*$.

As demonstrated in \citeref{Pess&Chan}, an analytical solution for the maximum growth rate and the critical wave number of the instability (named here $\pvar_{c}^{\nu}$) can be derived for visco-resistive disks. For vanishing resistivity, however, while the critical wave number becomes that of the ideal case $\pvar_{c}^{\nu}=\pvar_{c}$, \ie it does not directly depend on viscosity (see Appendix A), the maximum growth rate can not be derived analytically and the solution: 
\begin{align}
y_M^\nu=(\bar{s}^2\delta)^{-1/4}\sqrt{4-\bar{s}^2\delta/(2\viscbar)}\;,\label{Eqt.y_M}
\end{align}
has to be intended as the exact solution for very large viscosity and as an upper limit for $\viscbar>1$. It is worth stressing that even if the critical wavenumber coincides with that found in the inviscid case, the viscosity damping makes the mode amplitude for $\pvar\lesssim\pvar^\nu_{c}$ always negligible. Clearly, the instability is not only suppressed in amplitude by the vanishing $\delta$ value, but it is additionally damped for $\pvar_M^\nu<\pvar<\pvar_{c}$, where $\pvar_M^{\nu}$ corresponds the wavenumber associated to the maximum growth rate $y_M^\nu$. We conclude this Section underlining that the situation is reversed in the region where $\delta>\alpha$.

\section{Conclusions}
We developed a perturbation analysis of the stratified configuration concerning an incompressible plasma disk, by using the magnetic-flux function as the basic dynamical variable. We considered the background as associated to a purely differentially rotating inviscid stratified profile and we explicitly imposed the corotation theorem, \ie the dependence of the disk angular velocity on the unperturbed magnetic-flux function. We emphasize how, including viscosity on the background would not affect the validity of such a theorem, because the electron force balance equation is not influenced by viscosity.

We first analyzed the case of ideal perturbations to the background and then we included viscous effects. We demonstrated, in both these cases, the emergence of a vertical cut-off on the MRI in the positive $z$-axis, over a critical height and for a sufficiently large wavenumber (the same in the viscid and inviscid cases). However, the damping is much more marked in the presence of viscosity, since the growth rate is significantly suppressed already before the critical height and wavenumber. Such an asymmetry of the MRI can have significant consequences on the transport features of the plasma disk, especially when we recall that MRI is the only reliable mechanism to induce the necessary turbulence postulated by the Shakura idea for the accretion mechanism \cite{B&H1998,Bisno}. By other words, when the induced effective viscosity, responsible for a non-zero infalling velocity, is sufficiently relevant to influence the perturbation dynamics, the vertical asymmetry of the MRI (which we demonstrated to be already present in the ideal case) is enhanced so much that the accretion process can take place efficiently only on one side of the equatorial plane. On the other side, the MRI generates viscosity and it is, in turn, suppressed by its own product (actually the turbulent flow), so that we are led to think that the infalling velocity should be much weaker there. Clearly, this is just a qualitative statement, which requires further investigation to be applied to a real disk-like accretion structure. 

We conclude by stressing how the present analysis has a relevant astrophysical interest, since it concerns real plasma configurations accreting around compact objects (for a review on stellar accretion disks see \citeref{Bisno}, while for a discussion on the plasma stability within such systems, see \citeref{B&H1998}). Although the thin disk approximation succeeds in describing the basic features of many types of stellar accreting plasma, some structures are significantly thick and require a separate analysis, especially in view of their non-Keplerian differential rotation \cite{ogilvie97}. In this thick plasma configurations, the pressure gradients play an important role in fixing the steady equilibrium configuration and then in determining the behavior of linear perturbations. When the vertical pressure and mass density gradients are significantly stiff, the approximation of an angular frequency depending on the radial distance from the center only, appears rather rough. In fact, \eref{dhjbdslvbob} directly links the pressure and mass density vertical behaviors to the vertical variation of the angular frequency. The analysis here addressed applies just to systems of plasma arranged in such a way to be both thick and stratified disks. The choice of a dipole-like magnetic field is very reliable for accreting systems, whose background configuration does not react to the central object field, which in the disk region, \ie far enough from the star surface, is essentially described by its dipole component \cite{Bisno}.

We observe that a dipole-like magnetic field is essentially vertical in a thin disk configuration, almost coinciding with the equatorial plane of the accreting structure. Therefore, the most important deviation from the standard thin disk morphology is expected far enough from the equatorial plane, where the dipole-like field acquires a sufficiently large radial component. This is just what we observed in fixing a vertical quote for the MRI suppression both in the ideal and viscous cases. In this respect, the present study is reliably applicable to thick structures, having a sufficiently large vertical shear, like Advective Dominated Accretion Flows \cite{nara08}, but it can be also interesting for transient collapsing configurations, like Cataclysmic Variables \cite{smith06} (see also \citeref{pet13}).

Our study rises interesting questions concerning how the turbulence and accretion profiles \cite{balbus03} are deformed in view of the vertical shear. In particular, the suppression of MRI in the upper half plane of the configuration suggests that there the accretion mechanism can no longer rely on the effective viscosity due to turbulence and it would require alternative processes for the angular momentum transport.

\section{\appendixname\ A} 
Here, we are going to discuss the expression of the critical dimensionless wavenumber, $\pvar_c^\nu$, as it arises from the dispersion relation (\ref{Eqt.Dispersion}). In order to derive an equation for $\pvar_c^\nu$, we introduce the following dimensionless variables
\begin{align}
X = y + \viscbar\Gamma \;,\quad \Gamma = \pvar^2\;,\quad \bar{\varphi} =\varphi/\omega^2_0\;.
\end{align} 
Therefore, \eref{Eqt.Dispersion} can be rewritten in the equivalent form
\begin{align}
X^4 - 2\viscbar\delta X^3 + \left(2\delta + \viscbar^2\delta^2 + \bar{s}^2\right)X^2+\qquad\qquad\nonumber\\ 
-2\viscbar\delta\left(\delta + \bar{s}^2\right)X + \delta^2 + \beta\bar{\varphi}\delta + \viscbar^2\delta^2\bar{s}^2=0\;.
\end{align}
Clearly, at the critical point determined by the condition $y(\Gamma_{c})=0$, we have $X=\viscbar\Gamma_c(\pvar_c^\nu)$. Consequently, it is easy to verify that the following equation for $\Gamma_c$ holds: $(\Gamma_{c}+\bar{\varphi}\delta)\Gamma_{c}=0$, and the only non-trivial solution is $\pvar_c^\nu=\sqrt{-\delta \bar{\varphi}}$, which is equivalent to \eref{fhfdjksjdh}. The same result can be found in \citeref{Pess&Chan,Shak&Post} for vanishing magnetic resistivity.


\end{document}